\documentclass[preprint,12pt]{elsarticle}
\usepackage{hyperref}
\usepackage{amssymb}
\usepackage{amsfonts}
\usepackage{url}
\usepackage[normalem]{ulem}
\usepackage{amsmath}
\usepackage{xcolor}
\usepackage{graphicx}
\usepackage{subcaption}
\usepackage{bm}

\enlargethispage{\baselineskip}

\hypersetup{
    colorlinks=true,
    urlcolor=blue,
    citecolor=blue,
    linkcolor=blue,
    menucolor=blue,
    anchorcolor=blue,
    filecolor=blue
}

\journal{Physics of the Dark Universe}

\begin{document}
\begin{frontmatter}

\title{Compatibility of JWST results with exotic halos}

\author[a,b]{Fabio Iocco}
\ead{fabio.iocco.unina@gmail.com}
\author[c,d]{Luca Visinelli\corref{cor1}}
\ead{luca.visinelli@sjtu.edu.cn}
\cortext[cor1]{Corresponding author}

\address[a]{Universit\'a degli Studi di Napoli ``Federico II'',
                Via Cintia 26, 80126 Napoli, Italy}
\address[b]{INFN sezione di Napoli, Via Cintia 1, 80126 Napoli, Italy}
\address[c]{Tsung-Dao Lee Institute (TDLI), 520 Shengrong Road, 201210 Shanghai, China}
\address[d]{School of Physics and Astronomy, Shanghai Jiao Tong University, 800 Dongchuan Road, 200240 Shanghai, China}
                
\begin{abstract}
The James Webb Space Telescope (JWST) is unveiling astounding results about the first few hundred million years of life of the Universe, delivering images of galaxies at very high redshifts. Here, we develop a UV luminosity function model for high-redshift galaxies, considering parameters such as the stellar formation rate, dust extinction, and halo mass function. Calibration of this luminosity function model using UV luminosity data at redshifts $z = 4\textrm{-}7$ yields optimal parameter values. Testing the model against data at higher redshifts reveals successful accommodation of the data at $z = 8\textrm{-}9$, but challenges emerge at $z \simeq 13$. Our findings suggest a negligible role of dust extinction at the highest redshifts, prompting a modification of the stellar formation rate to incorporate a larger fraction of luminous objects per massive halo, consistently with similar recent studies. This effect could be attributed to mundane explanations such as unknown evolution of standard astrophysics at high redshift or to the existence of exotic objects at high redshift.
We comment on this latter possibility.
\end{abstract}

\end{frontmatter}

\section{Introduction}
\label{sec:introduction}

The James Webb Space Telescope (JWST) has begun the release of novel data on the cosmo, such as a population of massive galaxies with stellar masses $\gtrsim 10^{10},M_\odot$ at redshifts $z = 7.4\textrm{-}9.1$~\cite{2023Natur.616..266L} and UV-bright galaxies at even higher redshifts $z \gtrsim 10$~\cite{2022ApJ...938L..15C, 2022ApJ...940L..55F,2022ApJ...940L..14N,2023MNRAS.519.1201A,2023MNRAS.518.6011D, 2023MNRAS.519.3064F,2023ApJS..265....5H, 2023ApJ...946L..35M,2023ApJ...946L..13F,2023ApJ...942L..27S} that take on from the earlier work of the Hubble Space Telescope (HST)~\cite{Monna:2013eia, 2016ApJ...819..129O, Zheng:2017ebv}.

Data collected in the first year of operation already suggest that the Universe is teeming with young galaxies a few hundred million years after the Big Bang, to a somewhat surprising degree when compared with the expectations based on the predictions of the standard cosmological model. In fact, the early data released by the JWST seem to suggest both an excess of luminous objects, and an overabundance of the most massive ones --especially those at the highest detectable redshifts-- compared with earlier expectations based on the standard lore of structure formation~\cite{Gardner:2006ky, Haslbauer:2022vnq, 2023MNRAS.519.1201A,2022ApJ...940L..55F, 2023ApJS..265....5H,2023ApJ...955...13B, 2024ApJ...960...56H, 2023Natur.616..266L, 2023ApJ...942L...9Y, 2023MNRAS.518.4755A}.

The debate whether such an excess is truly persistent is far from being settled, see e.g.\ ref.~\cite{2023OJAp....6E..47M}, and the community has tried to address the topic with different methods~\cite{2023arXiv230507049S, 2023MNRAS.518.2511L, Boylan-Kolchin:2022kae,Chen:2023ugq,2023OJAp....6E..47M,2023MNRAS.524.2594K, 2023MNRAS.523.3201D, 2023MNRAS.521.3201N, Passaglia:2021jla, Biagetti:2022ode,Liu:2022okz,Liu:2022bvr, Hutsi:2022fzw, Batista:2017lwf, 2021MNRAS.504..769K, Adil:2023ara, Parashari:2023cui, Menci:2024rbq}. Collectively, these can be summarised into four main categories: {\it i}) a mundane explanation such as instrumental calibration; {\it ii}) astrophysical unknowns from dust absorption effects; {\it iii}) astrophysical unknowns from baryon (gas and dust) behavior: changes at very low or null metallicities would impact the physics of galaxy formation; {\it iv}) explanations invoking a population of high $z$ heavy compact objects. This last possibility has recently emerged as a compelling solution and a tool to test the standard cosmological model~\cite{Zurek:2006sy, Hutsi:2022fzw, Liu:2022okz, Liu:2022bvr, Bird:2023pkr, Huang:2023chx, Huang:2023mwy}. Examples include the standard picture of stellar evolution involving a first population of stars (PopIII) formed at redshifts $z\sim 10-20$ from the collapse of primordial zero metallicity clouds of molecular hydrogen~\cite{Hirano:2013lba}, and a population of first stars possibly powered by the annihilation of dark matter, either by accretion during its proto-stellar phases~\cite{Spolyar:2007qv}, or by capture once the actual stellar object is formed~\cite{Iocco:2008xb}.

In this work, we investigate the connection between the UV luminosity function model for high-redshift galaxies at different redshifts by testing the validity of a model against luminosity data. We construct a standard Press--Schechter halo density function and then populate this standard population of halos with luminosities obtained both in standard and exotic scenarios of star formation.

We confirm the mismatch between data and theoretical expectations when purely standard astrophysics is adopted and extrapolated at high redshift. We explore the possibility that such mismatch may be caused by some type of exotic object. Our approach is as follows: in section~\ref{sec:method}, we describe the model for the UV luminosity function and the magnitude from an individual object adopted to populate the standard halos, including dust extinction. In section~\ref{sec:ModCal}, we explain the methods used to calibrate the model against data at redshifts $z=4\textrm{-}7$ and present the best-fit parameters obtained with the chosen priors. Anticipating our results --presented in section~\ref{sec:res}-- we find that a population of ``standard'' halos at redshift $z \gtrsim 13$, with a higher star formation efficiency with respect to lower redshifts and no dust extinction, can explain the faint end of JWST observations. As we argue in our discussion presented in section~\ref{sec:disc}, such very luminous halos are compatible with enhanced star formation in very high redshift halos, accreting black holes, and some models of dark matter powered stars, i.e.\ {\tt dark stars}~\cite{Spolyar:2007qv}. None of the above alone seems however able to explain the higher luminosity end of the observations. A summary is provided in section~\ref{sec:summary}.

\section{Formalism and methodology}
\label{sec:method}

In this section, we describe the theoretical framework adopted throughout the paper. Firstly, we develop a Press-Schechter formalism to obtain the halo distribution function (number of halos per mass bin per redshift), as explained in subsection~\ref{sec:PS}. This is entirely consistent with the standard cosmological model. Only at the end of this section, we will populate these halos with a specific UV luminosity, tailored for a heavy stellar mass function, thereby customizing the halo luminosity function for the early Universe. We will explicitly state when this customization occurs.

In the second subsection~\ref{sec:MUV}, we illustrate the formalism used to compute the magnitude expected for individual objects, particularly those at high redshift. Later in the paper, we discuss this formalism in relation to exotic objects that potentially existed at high redshift. We will explicitly state when these objects are introduced in our analysis.

Setting our formalism, we find it useful to remind here the relation by Oke and Gunn~\cite{Oke:1983nt}\footnote{The paper contains a sign mistake in its most important equation.} giving the relative magnitude associated with the differential flux $F(\lambda, z)$ as
\begin{eqnarray}
	m_{\rm AB} &=& - 48.6 - 2.5\log_{10}\left(\frac{F(\lambda, z)}{{\rm erg \,s^{-1}\,cm^{-2}\,Hz^{-1}}}\right)\,,\\
	F(\lambda, z) &=& \frac{1}{4\pi d_L^2}\,(1+z) L_{\nu'}(\lambda')\,,
\end{eqnarray}
where $\lambda = (1+z)\lambda'$ is the wavelength observed today, redshifted with respect to the wavelength $\lambda'$ emitted at redshift $z$. Here, $L_{\nu'}(\lambda')$ is the flux emitted at source and $d_L$ is the luminosity distance at redshift $z$. The UV absolute magnitude $M_{\rm UV}$ is then obtained by combining the expressions above as
\begin{equation}
	\label{eq:magnitude}
	M_{\rm UV} = m_{\rm AB} - 5\log_{10}\left(\frac{d_L}{\rm10\,pc}\right) + 2.5\log_{10}(1+z) = 34.13-2.5\log_{10}\left(\frac{L_{\nu'}}{{\rm W\,Hz^{-1}}}\right)\,.
\end{equation}
Along with the absolute magnitude, a second measurement characterizing a distant object is the UV luminosity function $\Phi_{\rm UV}$ that tracks the emission and abundance of high-redshift galaxies. We define the UV luminosity function in terms of the halo mass function (HMF) introduced in the following section.

\subsection{Halo distribution function}
\label{sec:PS}

We consider the Press-Schechter formalism in a standard cosmological scenario, see e.g.\ ref.~\cite{Boylan-Kolchin:2022kae}. The differential halo mass function, corresponding to the number of halos per unit log mass per unit comoving volume, can be written as
\begin{equation}
    \label{eq:halomassfunction}
	\frac{{\rm d}n}{{\rm d}\,\ln M_h} = f(\xi)\,\rho_{m,0}\,\left|\frac{{\rm d}\,\ln\xi}{{\rm d}\,M_h}\right|\,,
\end{equation}
where $\rho_{m,0}$ is the average mass density today, $\xi \equiv \delta_c / \sigma$ with $\delta_c = 1.686$ is the critical density at collapse for a pressureless fluid, and $f(\xi)$ gives the halo mass function. The root mean square (rms) of linear density perturbations $\sigma$ for the enclosed mass $M_h$ is defined by considering the overdensity smoothed over the mass $M_h$ enclosed within the radius $R$ defined as
\begin{equation}
    M_h = \frac{4\pi}{3}\bar\rho_0R^3\,,
\end{equation}
where $\bar\rho_0$ is the mean density of the Universe at present time. This expression gives the relation $R = R(M_h)$ and the mass and redshift dependent rms density perturbations $\sigma(M_h, z) = D(z)\sigma(R(M_h))$, with the growth function
\begin{equation}
	D(z) \propto H(z)\int_z^{+\infty} {\rm d}z'\,\frac{1+z'}{[H(z')]^3}\,,
\end{equation}
normalized so that $D(0) = 1$. The variance of the density fluctuations smoothed over a region of radius $R$ is~\cite{Percival:2001nv}
\begin{equation}
    \label{eq:sigma}
	\sigma^2(R) = \int_0^{+\infty} \frac{{\rm d}k}{k} \Delta^2(k) \hat{W}^2(k, R)\,,
\end{equation}
in terms of the window function $\hat{W}(k, R)$ and of the dimensionless linear matter power spectrum $\Delta^2(k)$. More in detail, the window function associated with a top-hat filtering is
\begin{equation}
    \hat{W}(k, R) = \frac{3}{(kR)^3}\left[\cos(kR) - kR\sin (kR)\right]\,,
\end{equation}
while the power spectrum can be cast in terms of the primordial power spectrum $P_\zeta(k) \propto k^{n}$ and the matter transfer function $T(k)$ as
\begin{equation}
    \Delta^2(k) \equiv \frac{k^3}{2\pi^2}P_\zeta(k)\,T^2(k)\,.
\end{equation}
Here, the transfer function is provided through a fit to the numerical results from the Anisotropies in the Microwave Background (CAMB) code~\cite{Lewis:1999bs}.\footnote{Alternatively, an analytical fit is provided in ref.~\cite{Bardeen:1985tr}.}

The distribution in eq.~\eqref{eq:halomassfunction} can be evaluated using the code {\tt TheHaloMod} (previously {\tt HMFcalc})~\cite{Murray:2013qza, Murray:2020dcd}, which is archived on GitHub.\footnote{\href{https://github.com/halomod/TheHaloMod}{https://github.com/halomod/TheHaloMod}.} In the code, the function $f(\xi)$ describing the halo mass function is obtained from the numerical results in ref.~\cite{2010ApJ...724..878T}. We normalize the values in the code according to the analysis for the TT,TE,EE+lowE+lensing data by the Planck collaboration~\cite{Planck:2018vyg} and set $\sigma_8 = 0.8111$ for $R = 8\,h^{-1}{\rm\,Mpc}^{-1}$, the power spectrum index $n = 0.9646$ at the reference scale $k_0 = 0.05{\rm\,Mpc^{-1}}$, the matter density parameter $\Omega_m = 0.3153$, and we set the Hubble constant to $H_0 = 67.8{\rm\,km\,s^{-1}\,Mpc^{-1}}$. As a check, we have reproduced the Press-Schechter mass function and the rms density perturbation at $z = 0$ from ref.~\cite{dessert:online}. We remark that, until now, all equations are purely related to the Press-Schechter formalism, so we do not depend from any specific model parameter.

\subsection{Luminosity function}
\label{sec:MUV}

We estimate the specific UV luminosity per unit frequency appearing in eq.~\eqref{eq:luminosityfunct} as~\cite{Madau:2014bja}
\begin{equation}
    \label{eq:luminosityaccr}
    L_\nu = \eta_{\rm UV}\, \dot M_*\,,
\end{equation}
where the constant $\eta_{\rm UV}$ depends on the mass function and stellar population~\cite{Salpeter:1955it}. Here, we fix $\eta_{\rm UV} = 3.57 \times 10^{21}{\rm\,W\,Hz^{-1}}/(M_\odot{\rm\,yr^{-1}})$, corresponding to the value obtained for a zero-metallicity top-heavy initial mass function at 1500\,\AA~\cite{2022ApJ...938L..10I} for stellar masses in the range $M_*=(50\textrm{-}500)\,M_\odot$~\cite{Salpeter:1955it, 2022ApJ...938L..10I}.
The star formation rate depends on the halo accretion rate in terms of the stellar efficiency $f_h$ as~\cite{2022ApJ...938L..10I}
\begin{equation}
    \label{eq:deffstar}
    \dot M_* = f_h(M_h) \,\dot M_h\,.
\end{equation}
The function $f_h(M_h)$ defines the accretion model is here parameterized as
\begin{equation}
    \label{eq:SFR}
    f_h = \frac{f_*}{1+(M_*/M_h)^\gamma}\,,
\end{equation}
where we assess the parameters $(f_*, \gamma, M_*)$ through a numerical fit as described below in section~\ref{sec:res}. We do not include a dependence on redshift as it is yet unclear whether such a dependence is actually relevant at high redshifts~\cite{2022ApJ...929....1H}.

For the halo accretion rate, we use the identity
\begin{equation}
    \label{eq:dotMh}
    \dot M_h = -(1+z)\,H(z)\,\frac{{\rm d}M_h}{{\rm d}z}\,,
\end{equation}
where the last term is found within the extended Press-Schechter framework as~\cite{Neistein:2006ak, Correa:2014xma}
\begin{equation}
    \frac{{\rm d}M_h}{{\rm d}z} = \sqrt{\frac{2}{\pi}}\,\frac{M_h}{\sqrt{\sigma^2(M_h/q) - \sigma^2(M_h)}}\,\frac{\delta_c}{D^2(z)}\,\frac{{\rm d}D(z)}{{\rm d}z}\,.
\end{equation}
Here, the parameter $q$ is calibrated over numerical simulations, and we fix it here as $q = 2.2$. In the last step, we have defined the variance of the distribution at $z=0$ as $\sigma^2(M_h) \equiv \sigma^2(M_h, 0)$. 
Overall, the luminosity in eq.~\eqref{eq:luminosityaccr} is
\begin{equation}
    L_{\nu'} = -\eta_{\rm UV}\,f_h(M_h)\,\sqrt{\frac{2}{\pi}}\,\frac{\delta_c\,M_h}{\sqrt{\sigma^2(M_h/q) - \sigma^2(M_h)}} \,\frac{(1+z)\,H(z)}{D^2(z)}\,\frac{{\rm d}D(z)}{{\rm d}z}\,,
\end{equation}
which correctly reproduces the results $L_{\nu'} \propto (1+z)^{5/2}$ in the limit of large redshifts $z \gtrsim 1$~\cite{Dekel:2013uaa}. For this last result, we have approximated the growth function as $D(z) \approx A/(1+z)$ for $z \gtrsim 1$, where we find $A \approx 1.27$ with the parameter chosen. The derivative of the Oke-Gunn relation in eq.~\eqref{eq:magnitude} is
\begin{equation}
    \label{eq:dMUVdMh}
    \frac{{\rm d} M_{\rm UV}}{{\rm d}\,\ln M_h} = \frac{2.5}{\ln(10)}\,\left(1 - \frac{\frac{1}{q}\sigma\left(\frac{M_h}{q}\right)\sigma_M\left(\frac{M_h}{q}\right) - \sigma\left(M_h\right)\sigma_M\left(M_h\right)}{\sigma^2\left(\frac{M_h}{q}\right) - \sigma^2\left(M_h\right)} \right) \approx \frac{2.5}{\ln(10)}\,,
\end{equation}
where we have defined $\sigma_M(M_h) \equiv {\rm d}\sigma(M_h) / {\rm d}\ln M_h$.\footnote{Although the derivation of eq.~\eqref{eq:dMUVdMh} is straightforward, see e.g.\ ref.~\cite{Mason:2022tiy}, to the best of our knowledge the inclusion of the sigmas derivatives has not appeared previously.} A concise expression for the luminosity function within the model is then
\begin{equation}
    \label{eq:luminosityfunct}
    \Phi_{\rm UV} \equiv \frac{{\rm d} n}{{\rm d}M_{\rm UV}} = \frac{{\rm d} n}{{\rm d}\,\ln M_h} \bigg/ \, \frac{{\rm d}M_{\rm UV}}{{\rm d}\,\ln M_h}\,,
\end{equation}
where the first term is the HMF in eq.~\eqref{eq:halomassfunction} that describes the number of galaxies within the mass range $[M_h, M_h + {\rm d}M_h]$ per unit volume, while the second term is the halo-galaxy connection that relates the mass and the luminosity of a specific halo. For a different approach in the literature see e.g.\ ref.~\cite{Wang:2023xmm}. Using the Oke-Gunn relation in eq.~\eqref{eq:magnitude} we finally obtain
\begin{equation}
	\label{eq:magnitude_luminosity}
	M_{\rm UV} = -12.25 -2.5\log_{10}\left(f_h(M_h)\right) -2.5\log_{10}\left(\frac{\dot M_h}{10^{-3}{\rm\, M_\odot\,yr^{-1}}}\right)\,.
\end{equation}
We further include the effect of dust, whose general effect is that of absorbing light in the ultraviolet and optical frequencies, which are then re-emitted in the infrared~\cite{Spitzer:1990ec, Witt:1992aa}. Here, dust is parameterized by employing a correction to the UV magnitude by the quantity $A_{\rm UV}$~\cite{Meurer:1999jj, 2018ApJ...853...56R}, with the empirical determination of the slope $\beta = \beta(M_{\rm UV}, z)$ for $z<8$~\cite{Bouwens:2013hxa} and at higher redshifts~\cite{2023MNRAS.520...14C, 2024MNRAS.531..997C}, using the modeling given in refs.~\cite{Trenti:2014hka, Mason:2015cna}.

\subsection{Model calibration}
\label{sec:ModCal}

\begin{figure}[ht]
    \centering
    \includegraphics[width=0.85\linewidth]{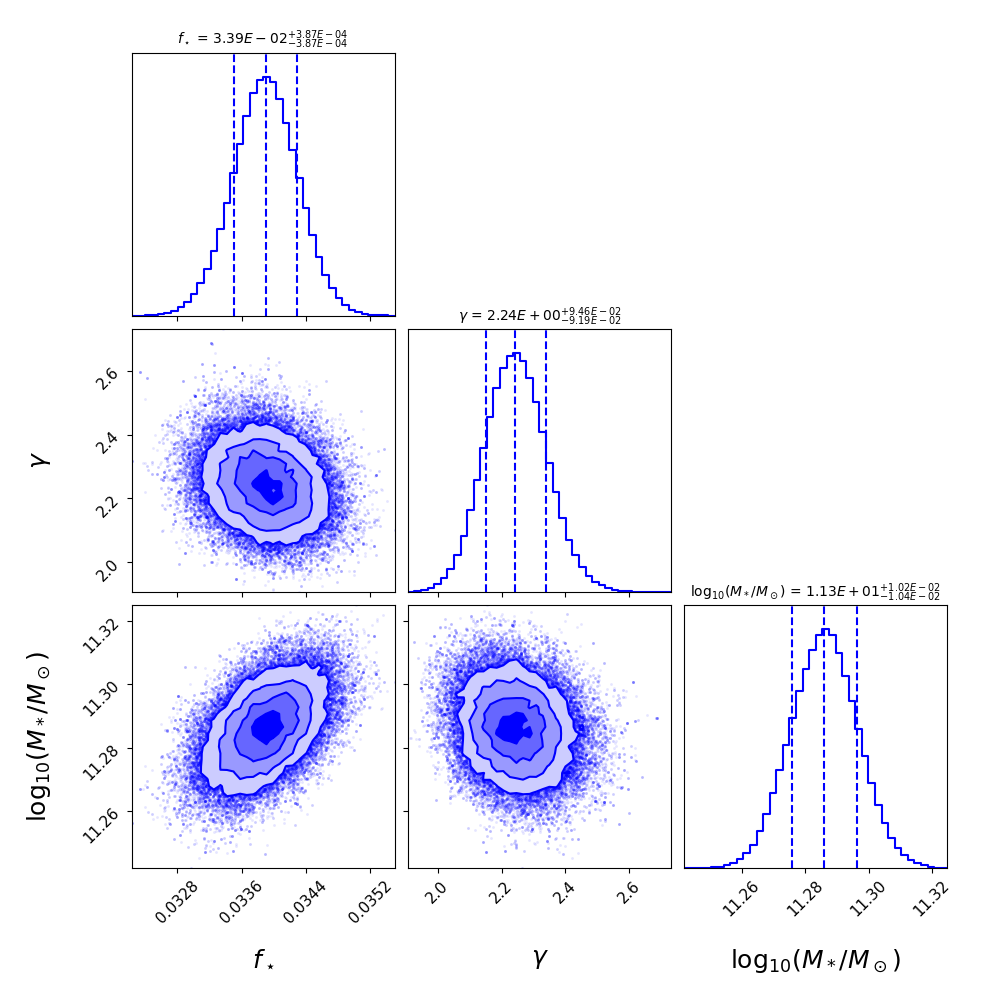}
    \caption{Posterior distributions of the parameters $(f_*, \gamma, M_*)$ defined in eq.~\eqref{eq:SFR} for the best-fit UV luminosity functions in figure~\ref{fig:UVLF}. The contours show the 1,2,3$\sigma$ credibility ranges and the values quoted correspond to the 1$\sigma$ level.}
    \label{fig:Phivsmag}
\end{figure}

We calibrate the model described above against the data at redshifts $z \sim 4-7$ obtained from refs.~\cite{McLure:2012fk, 2015ApJ...810...71F, 2021AJ....162...47B, 2022ApJS..259...20H}. The priors for the parameters are defined as in table~\ref{table:priors}. For the likelihood, we consider a split normal distribution, that allows asymmetry around the mode and yields the normal distribution in the special case of symmetry~\cite{doi:10.1080/01621459.1998.10474117}. Given the $i$-th data point $(M_i, \Phi_i)$ at redshift $z$, the calibration is performed by first obtaining the halo mass $M_h$ satisfying $M_{\rm UV}(M_h, z) = M_i$, then computing the difference $\delta_i \equiv \Phi_{\rm UV}(M_h, z) - \Phi_i$ and the corresponding likelihood. Table~\ref{table:priors} shows the priors assumed for the Monte Carlo analysis and the confidence level at 1$\sigma$ for the parameters considered. The results for the posterior and marginal distributions are reported in figure~\ref{fig:Phivsmag}.
\begin{table}
    \begin{center}
    \renewcommand{\arraystretch}{1.4}
    \begin{tabular}{|l@{\hspace{0.1 cm}}|l@{\hspace{0.1 cm}}|l@{\hspace{0.1 cm}}|l@{\hspace{0.1 cm}}|}
    \hline
    \textbf{Parameter}          & \textbf{Prior}       & \textbf{Best fit}         & \textbf{Calibration}\\
    \hline\hline
    $f_*$                       & Uniform on $[0.0\,,\,1.0]$  &  $3.39\times 10^{-2}$ & $(3.39\pm 0.04)\times 10^{-2}$\\
    $\log_{10}(M_*/M_\odot)$    & Uniform on $[6\,,\,15]$     & 11.29 & $11.29\pm 0.01$\\
    $\gamma$                     & Uniform on $[0.0\,,\,5.0]$ &  2.24 & $2.24\pm 0.01$\\
    \hline
    \end{tabular}
    \end{center}
    \caption{The table lists the parameters considered in the Monte Carlo analysis related to the model in eq.~\eqref{eq:SFR} assumed in this paper. Also shown is the range of the flat priors used in the analysis, the best fit values of the analysis and the results of the calibration at 68\% confidence level.}
    \label{table:priors}
\end{table}

\section{Results}
\label{sec:res}

\begin{figure}
    \centering
    \begin{subfigure}[t]{0.48\textwidth}
        \centering
        \includegraphics[width=\linewidth]{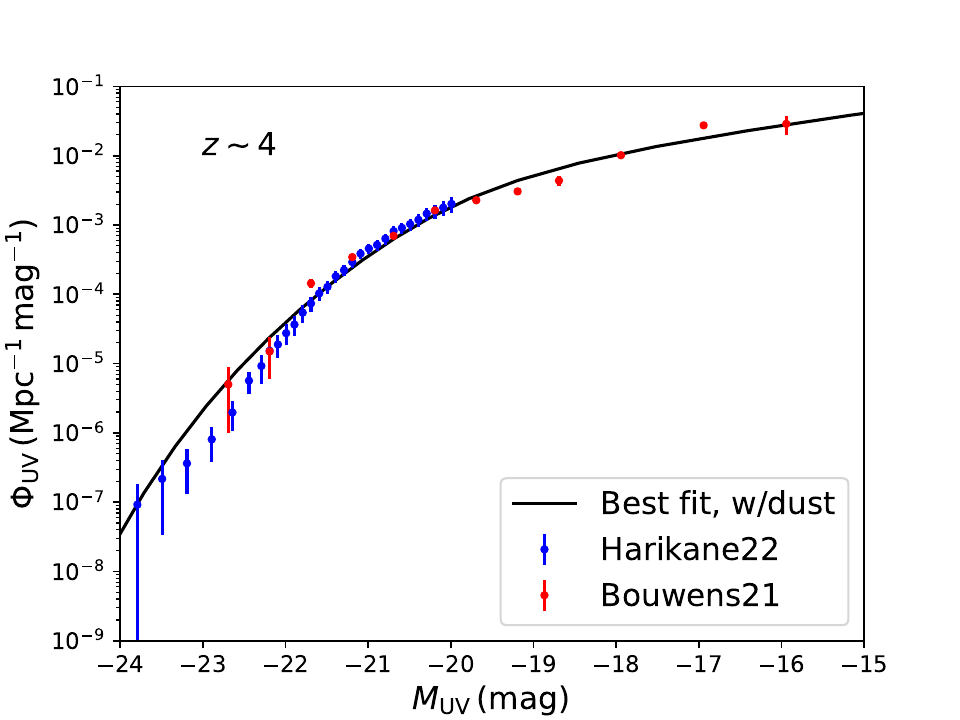} 
    \end{subfigure}
    \hfill
    \begin{subfigure}[t]{0.48\textwidth}
        \centering
        \includegraphics[width=\linewidth]{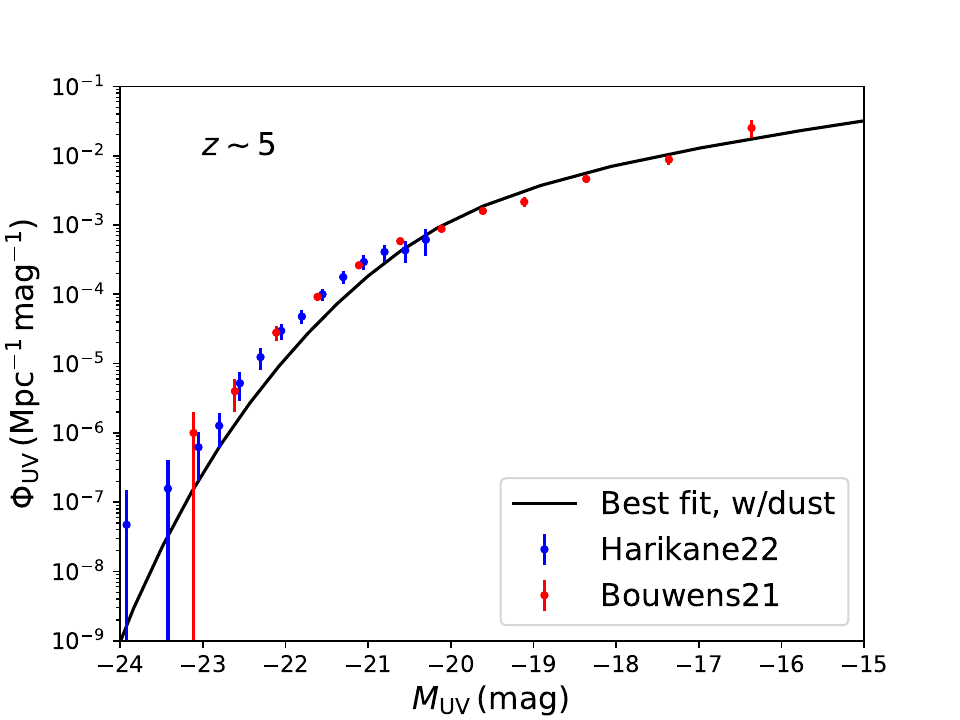} 
    \end{subfigure}
    \vspace{0.5cm}
    \begin{subfigure}[t]{0.48\textwidth}
        \centering
        \includegraphics[width=\linewidth]{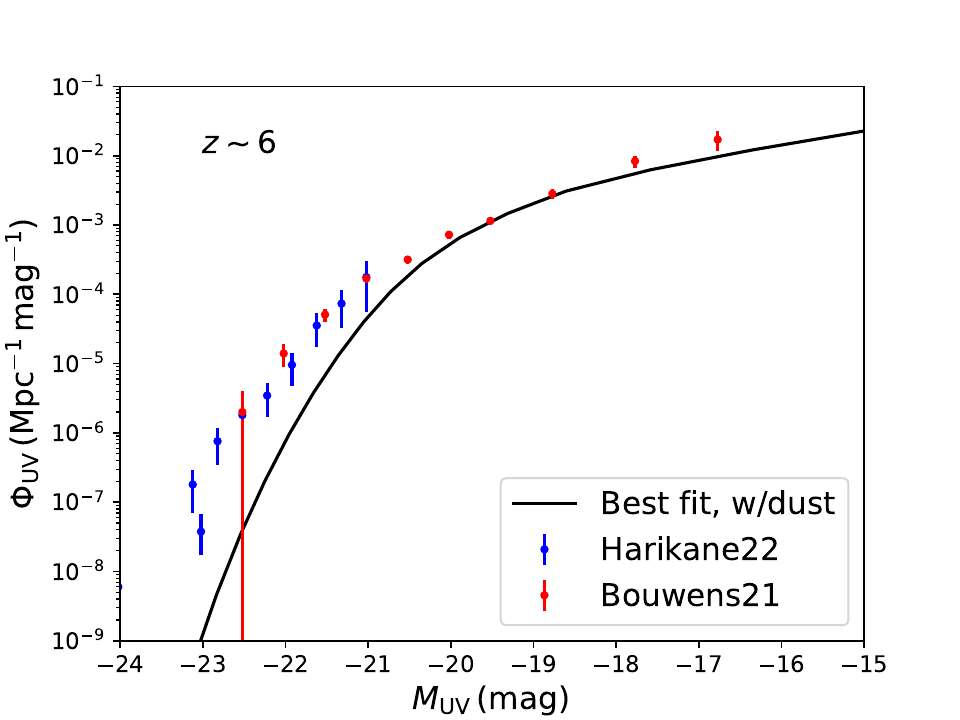} 
    \end{subfigure}
    \hfill
    \begin{subfigure}[t]{0.48\textwidth}
        \centering
        \includegraphics[width=\linewidth]{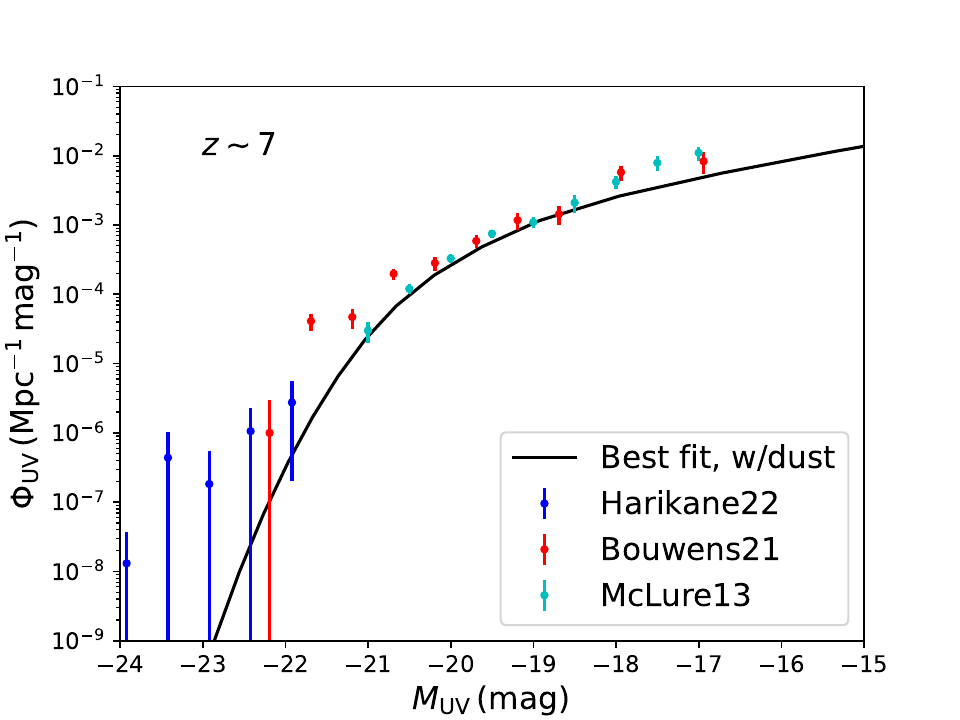} 
    \end{subfigure}
    \caption{The UV luminosity function $\Phi_{\rm UV}$ defined in eq.~\eqref{eq:luminosityfunct} with the best fit from the numerical analysis in table~\ref{table:priors} and the inclusion of dust extinction for the redshifts $z \sim 4-7$ (solid black lines). The result is compared against the data from Harikane22~\cite{2022ApJS..259...20H}, Bouwens21~\cite{2021AJ....162...47B}, and McLure13~\cite{McLure:2012fk}.}
    \label{fig:UVLF}
\end{figure}

We show here the predictions of our model as described and calibrated in the previous section. Figure~\ref{fig:UVLF} shows the luminosity function $\Phi_{\rm UV}$ obtained from the model in eq.~\eqref{eq:luminosityfunct} as a function of the Oke-Gunn magnitude $M_{\rm UV}$ in eq.~\eqref{eq:magnitude_luminosity}. The parameters are chosen according to the best fit obtained against the data used for the calibration, as expressed in table~\ref{table:priors}. 

As one can see, at redshifts $z\sim 6-7$ the luminosity of observed bright objects shows an excess compared with the model predictions, which could be attributed to the significant uncertainties in the data and a reduced weight for luminous objects in the fitting process. 

It is compelling to comment here on the choice of the spectral function. The halo density function at high redshift is sensible to the transfer function $T(k)$ and rms density $\sigma(R)$ adopted. For the main results presented until now, and in the following, we have adopted the transfer function and rms density as from CAMB prescriptions. If instead, one adopts a different value such as what is expressed in ref.~\cite{2022ApJ...938L..10I}, the halo mass function and henceforth the curves in the flux--magnitude plane change. We have checked that for a range of choices our conclusions with different transfer function $T(k)$ and rms density $\sigma(R)$ are not modified.

\begin{figure}
    \centering
    \begin{subfigure}[t]{0.48\textwidth}
        \centering
        \includegraphics[width=\linewidth]{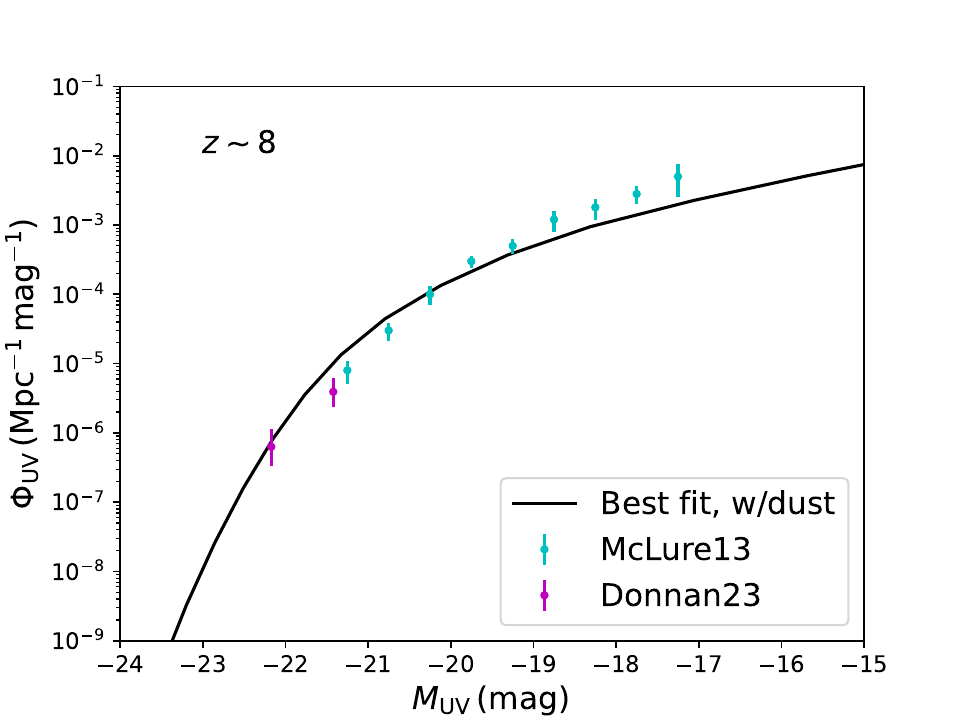} 
    \end{subfigure}
    \hfill
    \begin{subfigure}[t]{0.48\textwidth}
        \centering
        \includegraphics[width=\linewidth]{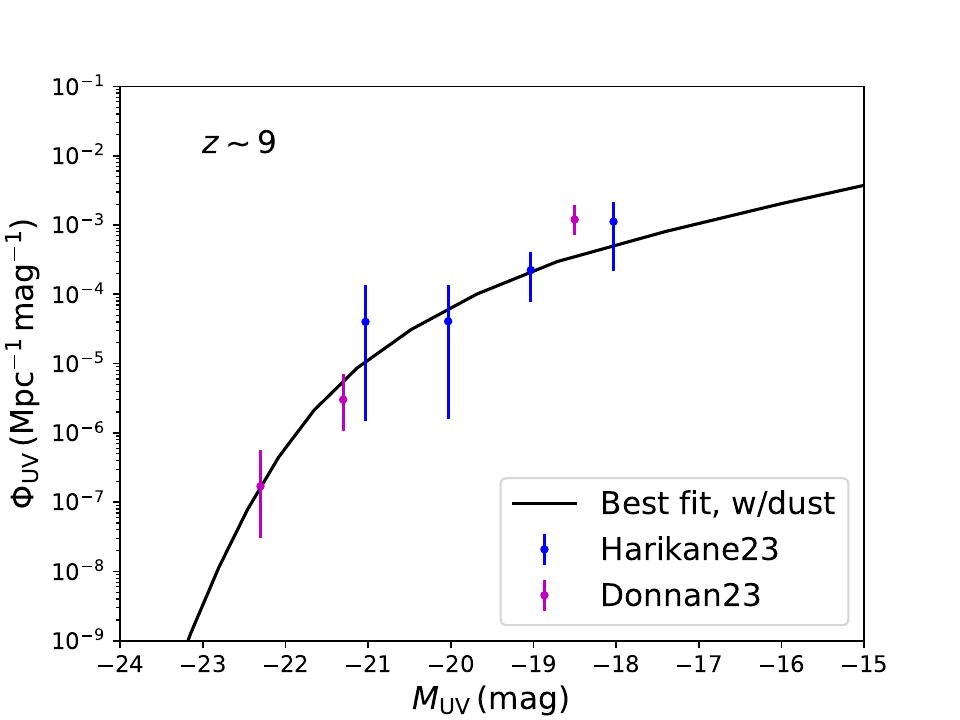} 
    \end{subfigure}
    \caption{The UV luminosity function $\Phi_{\rm UV}$ defined in eq.~\eqref{eq:luminosityfunct} with the best fit from the numerical analysis in table~\ref{table:priors} and the inclusion of dust extinction for the redshifts $z \sim 8\textrm{-}9$ (solid black lines). The result is compared against the data from McLure13~\cite{McLure:2012fk}, Harikane23~\cite{2023ApJS..265....5H}, and Donnan23~\cite{2023MNRAS.518.6011D}.}
    \label{fig:UVLFz89}
\end{figure}

As a test, in figure~\ref{fig:UVLFz89} the model is compared against the data at redshift $z\sim 8$ from refs.~\cite{McLure:2012fk, 2023MNRAS.518.6011D} and $z\sim9$ from refs.~\cite{2023MNRAS.518.6011D, 2023ApJS..265....5H}. At these high redshifts, a description of the luminosity function is required in terms of a halo fraction dependent on the halo mass, as well as the magnitude correction from dust extinction.

For these redshifts, the best-fit values of the parameters obtained in the Monte Carlo analysis lead to a good fit of the considered data.

\section{Discussion}
\label{sec:disc}

We now turn to discussing the bright objects reported in recent literature~\cite{2022ApJ...929....1H, 2023ApJS..265....5H, 2023MNRAS.518.6011D}. We consider the data at redshifts $z \sim 8-13$ from table~6 of ref.~\cite{2023MNRAS.518.6011D}, while data collected by JWST for $z=13$ are from refs.~\cite{2022ApJ...929....1H, 2023MNRAS.518.6011D}. To frame these results in the context of JWST observations, in the following we focus on the case $z = 13$.

\begin{figure}
    \centering
    \includegraphics[width=0.7\linewidth]{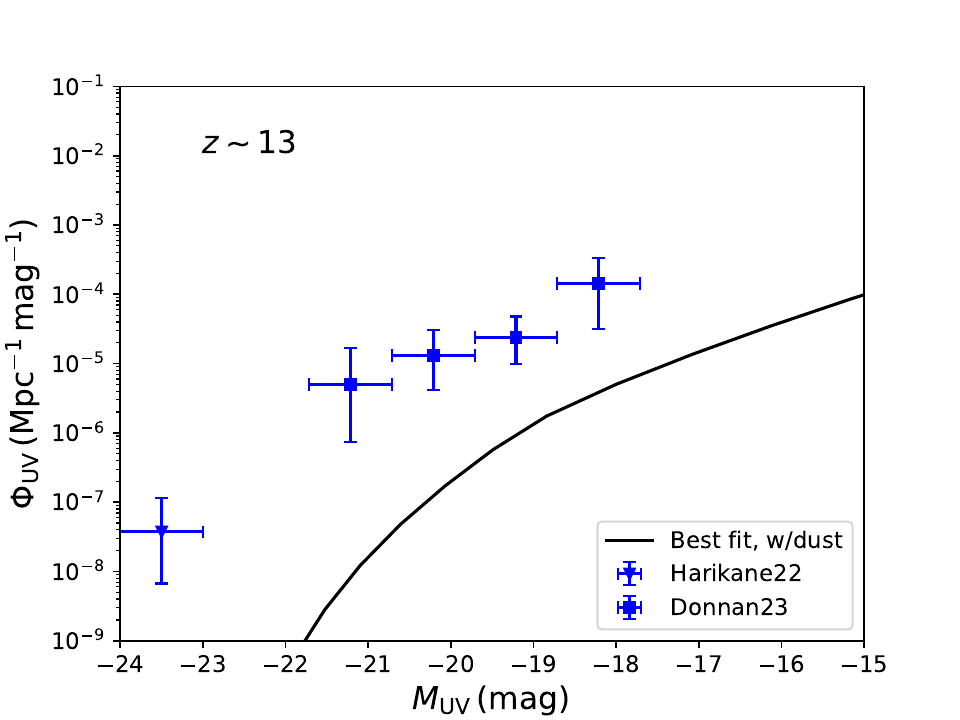}
     \caption{The UV luminosity function $\Phi_{\rm UV}$ defined in eq.~\eqref{eq:luminosityfunct} for the redshift $z \sim 13$, with the inclusion of dust extinction and the SFR parameters in eq.~\eqref{eq:SFR} taken from the best fit of the numerical analysis in table~\ref{table:priors} (solid black lines). The results are compared against the data from Harikane22~\cite{2022ApJ...929....1H} and Donnan23~\cite{2023MNRAS.518.6011D}.}
    \label{fig:Phivsmagz13a}
\end{figure}

The curves in the luminosity-magnitude plane of figure~\ref{fig:Phivsmagz13a} show what to expect from a standard population of halos powered by different combinations of star formation parameters. Higher mass halos naturally populate the higher luminosity and lower flux corner (bottom left), while smaller halos populate the lower luminosity and higher flux (top right), being also more numerous at any given redshift. The solid black line shows the result for the model obtained with the calibration in table~\ref{table:priors}, leading to values for the luminosity function which are generally lower than the data reported by orders of magnitude and a mismatch at more than 2$\sigma$. Thus, the model does not lead to a satisfactory match with data, which leads to questioning the assumptions made on the dust model and the validity of eq.~\eqref{eq:SFR}. 

\begin{figure}
    \centering
    \includegraphics[width=0.7\linewidth]{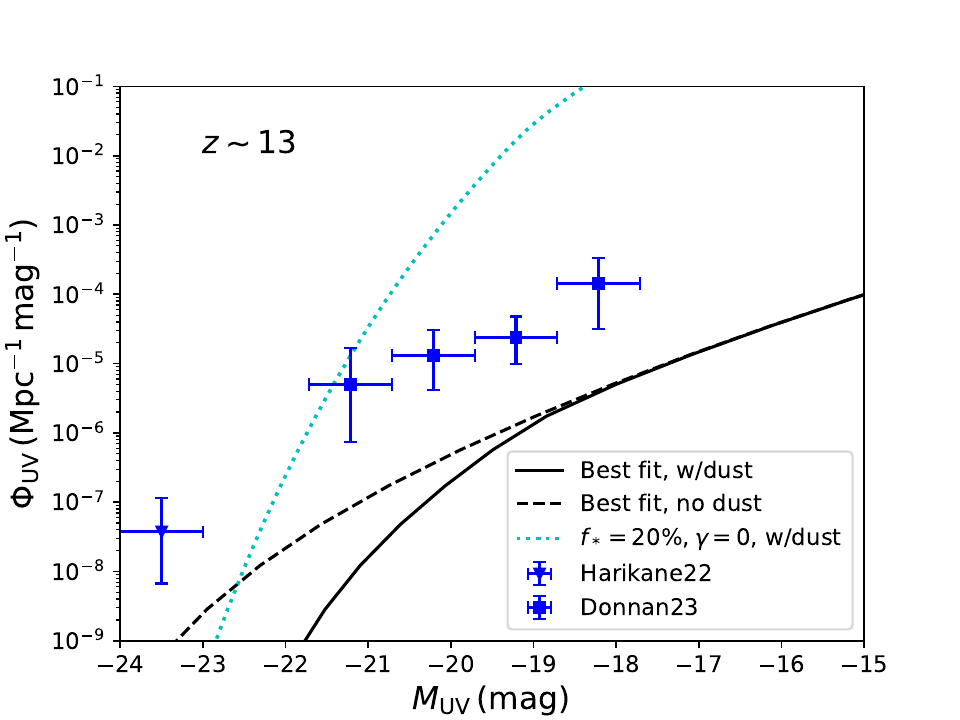}
     \caption{The UV luminosity function $\Phi_{\rm UV}$ defined in eq.~\eqref{eq:luminosityfunct} for the redshift $z \sim 13$, with the inclusion of dust extinction and the SFR parameters in eq.~\eqref{eq:SFR} taken from the best fit of the numerical analysis in table~\ref{table:priors} (solid black lines). Also shown are i) the function $\Phi_{\rm UV}$ without the inclusion of dust extinction and with the same best fit for the SFR parameters as in table~\ref{table:priors} (black dashed line), and ii) the function $\Phi_{\rm UV}$ with dust extinction and with the SFR parameters $f_* = 20\%$ and $\gamma = 0$ (cyan dotted line). The results are compared against the data from Harikane22~\cite{2022ApJ...929....1H} and Donnan23~\cite{2023MNRAS.518.6011D}.}
    \label{fig:Phivsmagz13b}
\end{figure}

Some attempts towards an understanding of the features of the UV luminosity function that is consistent with very high redshift data are explored in figure~\ref{fig:Phivsmagz13b}. For instance, dust extinction could play a less remarkable role at redshifts $z \gtrsim 10$, as underlined in recent observations by the JWST~\cite{2022ApJ...938L..15C, 2022ApJ...940L..55F, 2022ApJ...940L..14N, 2023MNRAS.519.1201A, 2023MNRAS.518.6011D, 2023MNRAS.519.3064F, 2023ApJS..265....5H, 2023ApJ...946L..35M, 2023MNRAS.522.3986F}. This is in sharp contrast with the dusty environments required to explain the stellar-forming galaxies at redshifts $z\sim 4\textrm{-}7$ observed by the Atacama Large Millimeter Array (ALMA)~\cite{2015Natur.519..327W, 2017MNRAS.466..138K, 2017ApJ...837L..21L, 2019PASJ...71...71H, 2019ApJ...874...27T}. Several solutions involving astrophysical processes have been proposed and can be tested with ALMA observations~\cite{2023MNRAS.520.2445Z}. While a conclusion might not yet be drawn, the results above stress that a complicated history of dust evolution occurred.

We find that eliminating the dust contribution alone is not a satisfactory choice as shown by the black dashed line, which consider the model with the best fit parameters from table~\ref{table:priors} and no contribution from dust extinction to the magnitude. Overall, the removal of dust extinction leads to the correct shape of the UV luminosity function although a mismatch of the SFR appears. A different approach consists in keeping the dust extinction while modifying the parameters that define the SFR function in eq.~\eqref{eq:SFR}. We attempt to increase the fraction of luminous objects per halos by setting the parameter $f_*=20\%$ with no halo mass dependence on the SFR, $\gamma = 0$ (dotted cyan line), thus leading to a constant stellar fraction $f_h = 10\%$. However, the shape and the slope of the resulting function does not match the observed data by several orders of magnitude.

\begin{figure}
    \centering
    \includegraphics[width=0.7\linewidth]{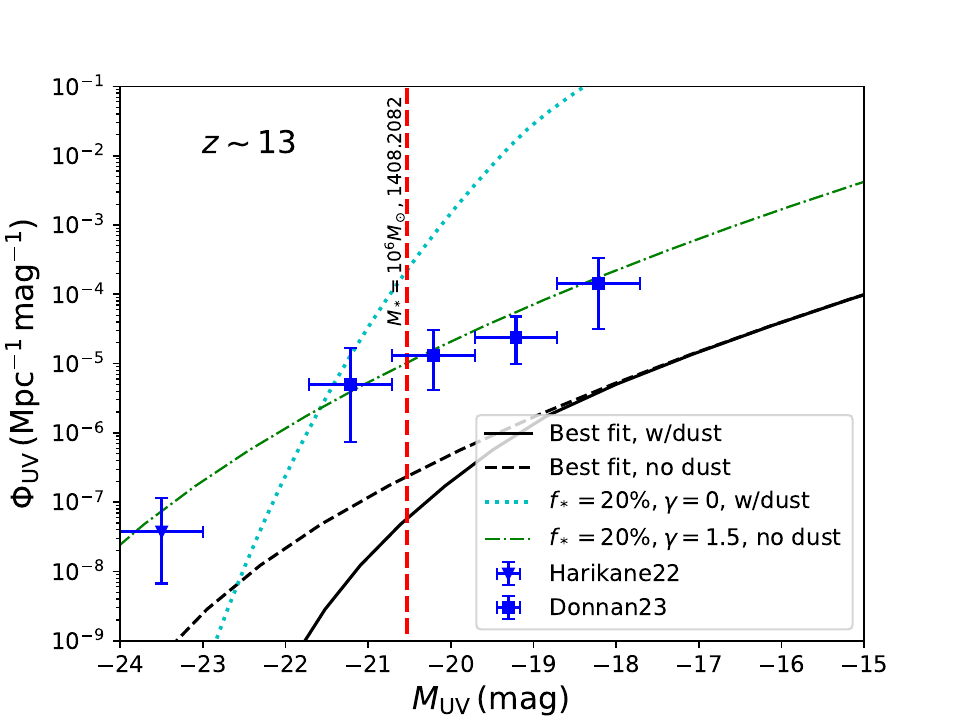}
    \caption{The UV luminosity function $\Phi_{\rm UV}$ defined in eq.~\eqref{eq:luminosityfunct} for the redshift $z \sim 13$, with the inclusion of dust extinction and the SFR parameters in eq.~\eqref{eq:SFR} taken from the best fit of the numerical analysis in table~\ref{table:priors} (solid black lines). Also shown are i) the function $\Phi_{\rm UV}$ without the inclusion of dust extinction and with the same best fit for the SFR parameters as in table~\ref{table:priors} (black dashed line), ii) the function $\Phi_{\rm UV}$ with dust extinction and with the SFR parameters $f_* = 20\%$ and $\gamma = 0$ (cyan dotted line), and iii) the function $\Phi_{\rm UV}$ without the inclusion of dust extinction and with the SFR parameters $f_* = 20\%$ and $\gamma=1.5$ (green dot-dashed line). The results are compared against the data from Harikane22~\cite{2022ApJ...929....1H} and Donnan23~\cite{2023MNRAS.518.6011D}. Red vertical line is a {\tt dark star} model in ref.~\cite{Rindler-Daller:2014uja}.}
    \label{fig:Phivsmagz13}
\end{figure}

We do however find that the results from the JWST observations can be explained by a model with negligible dust and a suitable stellar efficiency $f_*$ with a considerable contribution from stars $f_*=20\%$ and a lower slope $\gamma \simeq 1.5$ compared to the best fit results. This is summarized in figure~\ref{fig:Phivsmagz13} (green dot-dashed line), where we also show the functions discussed before in figures~\ref{fig:Phivsmagz13a} and~\ref{fig:Phivsmagz13b}. We stress here that all results obtained so far rely on standard scenarios, and a choice of non--exotic parameters. In the following section we offer a discussion of the possible physics of the halo necessary to achieve such values such as those described in ref.~\cite{Rindler-Daller:2014uja} (vertical red dashed line). Similar conclusions are drawn in ref.~\cite{2024A&A...684A.207F} using a different assumption for the SFR and a different feedback modeling.

\subsection{Exotic objects}
\label{sec:DS}
We speculate here on the nature of the objects that populate the massive halos at such high redshifts. The data points sketched in figure~\ref{fig:Phivsmagz13} could be resulting from the presence of exotic objects at high redshift, besides hydrogen--burning Population III stars.

Dark matter concentrated in the center of the halo by the cloud collapse could -through an annihilation mechanism--  affect the star formation phases in primordial halos~\cite{Ascasibar:2006pf}, and potentially lead to a new phase of stellar evolution~\cite{Spolyar:2007qv}, an object often referred to as a {\tt dark star}.

Furthermore, once a primordial star is formed, it could be collecting the dark matter streaming through it via a mechanism driven by the scattering between the ordinary stellar material and the dark matter, which once captured inside the star could power it for longer times and altering its structure~\cite{Iocco:2008xb, Freese:2008ur, Iocco:2008rb, Yoon:2008km}. To avoid confusion between the two dark matter collection mechanisms --direct gravitationally driven accretion {\it vs} scatter mediated capture-- we will be explicit about the type of mechanism that accretes the dark matter within the object.


A large body of work has been devoted to the effects of dark matter over the formation and evolution of the first stars in the Universe, and we do not seek to review that luscious literature here. For this, see refs.~\cite{Freese:2015mta, Iocco:2012jt} and references therein. Despite all efforts and arguments raised, the studies focusing on the effects of dark matter on the formation and evolution of the first stars have not ruled out any of the possibilities above~\cite{Smith:2012ng, Zackrisson:2010jd}, and the door remains open to speculation on whether any still viable self-annihilating dark matter candidate will produce effects along the lines of those described above on the first generation of stars~\cite{Wu:2022wzw}.

Here we only seek to understand whether any of the high redshift mismatches, such as the existence of high-luminous objects highlighted in the previous section, can be explained by some of the exotic dark matter-powered objects described in the literature of the past decade, leaving further details to other more focused studies.\footnote{It is fitting to notice here, for instance, that the analysis in ref.~\cite{Ilie:2023zfv} has already identified one object --JADES-GS-z13-0, at redshift $z\approx 13$ with an intrinsic magnitude $M_{\rm UV} = -18.5\pm 0.2$~\cite{Robertson:2022gdk}-- as a massive dark star.}

In our own analysis here, we start with those direct dark matter accretion driven objects --usually referred to as {\tt dark stars}-- adopting the structure modeling of ref.~\cite{Rindler-Daller:2014uja}.\footnote{See also refs~\cite{Freese:2010re, Rindler-Daller:2020yqe} for other studies of dark star structures.} The vertical red dashed line in figure~\ref{fig:Phivsmagz13} shows the magnitude obtained from an individual massive dark star as described therein, namely $M = 10^6\,M_\odot$ hosted in a halo of mass $M_h = 10^8\,M_\odot$, with a temperature $T = 3.14\times 10^4\,$K and an accretion rate $\dot M_h = 10^{-1}\,M_\odot{\rm\,yr}^{-1}$. Here, the dark star spectrum is approximated by that of a black body, which is a reasonable assumption for frequencies $\nu \gtrsim 0.1\,\mu$m~\cite{2012MNRAS.422.2164I}. We do not provide here with a model for the luminosity function describing these objects, which could therefore lie at different locations on the vertical axis of figure~\ref{fig:Phivsmagz13}. One possible way to fix this would be to relate a single star to the UV function $\Phi_{\rm UV}$ with the assumption that only one object per halo is formed. Based on current knowledge, this methodology would lead to utterly inconclusive considerations on the flux. For this reason we have preferred to leave the flux coordinate undetermined in figure~\ref{fig:Phivsmagz13}.

On the opposite, objects in which dark matter is accreted via a scatter--driven capture mechanism can not achieve such high magnitudes
~\cite{Freese:2008ur, Yoon:2008km, Taoso:2008zz}, and also, in order to sustain its dark matter-burning phase, a star must occupy a region with very dense dark matter environments. This mechanism is difficult to be achieved continuously over cosmological timescales~\cite{Sivertsson:2010zm, Iocco:2010fxs, Iocco:2012jt}, thus making it less efficient in delivering observable effects at redshifts below $z\approx12$. 
The advance of future observations in detecting the specific spectral shape of a high redshift object will be capable of shedding light on this~\cite{Zhang:2023kfj}. The search can be eased by the amplified magnitude resulting from gravitational lensing at such high redshifts~\cite{2024MNRAS.533.2727Z}.

It is to be noted that the {\tt dark star} object reported in our figure~\ref{fig:Phivsmagz13} is the most extreme for which ref.~\cite{Rindler-Daller:2014uja} provides a detailed structure, resulting from the direct collapse of a total mass of baryons $M = 10^6\,M_\odot$ into a single object, hosted inside a halo of mass $M_h=10^8\,M_\odot$. It is improbable that a single object predicted in these models can achieve the higher luminosities that are required to interpret the high absolute magnitudes shown in figure~\ref{fig:Phivsmagz13}, especially when several orders of magnitude in mass are required to explain the left-end observations. Consequently, it seems unlikely that individual stellar objects powered by dark matter annihilation can account for the entirety of the JWST observations, and more study on how massive halos could host multiple exotic stars is required.

\section{Summary and conclusions}
\label{sec:summary}

We have confirmed the claims in the recent literature that the results reported by the James Webb Space Telescope (JWST) at high redshift are not easily accommodated within a Press-Schechter model for the UV luminosity function and dust-corrected intrinsic magnitude that fits the data at lower redshifts. To assess this, we constructed a model for the UV luminosity function calibrated with data reported at redshifts $z = 4\textrm{-}7$, with a stellar formation rate (SFR) that depends on the halo mass as in eq.~\eqref{eq:SFR}. The magnitude associated with a given halo from eq.~\eqref{eq:magnitude_luminosity} is corrected by a term $A_{\rm UV}$ to account for dust extinction.

While the best fit parameters of the model explain the data at higher redshifts $z = 8\textrm{-}9$, the most recent data at even higher redshifts are harder to reconcile with this method. As an example, we have compared the predictions for the UV luminosity from our best fit model with the data at redshift $z\sim 13$ from refs.~\cite{2022ApJ...929....1H, 2023MNRAS.518.6011D}. The results are summarized in figure~\ref{fig:Phivsmagz13}, where we generally obtain that dust extinction is not required at such high redshifts, consistently with recent work on the topic. Moreover, the SFR should be modified to include a larger fraction of luminous objects per massive halo $f_*$, with a milder dependence onto the halo mass distribution $\gamma$.

In summary, existing models of exotic high redshift objects --dark matter powered structures often referred to as {\tt dark stars}-- are capable of explaining the data at lower absolute magnitudes. However, these models are likely to encounter challenges when explaining the behavior at the highest luminosity end of the observations, especially when interpreted in terms of individual objects. 

\section*{Acknowledgements}
We thank A.~Ferrara and K.~Freese for a careful read of a preliminary version of the draft, as well as for useful comments that have 
improved the manuscript.
F.I.\ is partially supported by the research grant No.\ 2022E2J4RK ``PANTHEON: Perspectives in Astroparticle and Neutrino THEory with Old and New messengers'' under the program PRIN 2022 funded by the Italian Ministero dell’Universit\`a e della Ricerca (MUR) and by the European Union – Next Generation EU.
F.I.\ acknowledges the hospitality of the Tsung-Dao Lee Institute (TDLI) in Shanghai (China), and Universit\`a di Ferrara (Italy) during the final phases of this work, funded by the TAsP Iniziativa Specifica of INFN.
L.V.\ acknowledges support by the National Natural Science Foundation of China (NSFC) through the grant No.\ 12350610240, as well as hospitality by the Istituto Nazionale di Fisica Nucleare (INFN) section of Napoli (Italy), the Galileo Galilei Institute for Theoretical Physics in Firenze (Italy), the INFN section of Ferrara (Italy), the University of Texas at Austin (USA), and the INFN Frascati National Laboratories near Roma (Italy) throughout the completion of this work.

\clearpage

\bibliographystyle{JHEP}
\bibliography{JWST.bib}

\end{document}